\documentclass{iopart}

\usepackage{graphicx,bbm,amssymb,amsopn}

\usepackage[english]{babel} 
\usepackage{dsfont}

\begin{document}

\title[Transport through quantum dot coupled to superconducting leads]{Markovian kinetic equation approach to electron transport through quantum dot coupled to superconducting leads}

\author{Daniel S. Kosov${}^{1,2}$, Toma\v z Prosen${}^3$, and Bojan \v{Z}unkovi\v{c}${}^{3}$}

\address{
${}^1$ School of Engineering and Physical Sciences,
James Cook University,
Townsville, QLD, 4811, Australia\\
${}^2$ Department of Physics, 
Universit\'e Libre de Bruxelles, Campus Plaine, CP 231, Blvd du Triomphe, B-1050 Brussels, Belgium\\
${}^3$Department of Physics, Faculty of Mathematics and Physics, University of Ljubljana, Jadranska 19, SI-1000 Ljubljana, Slovenia}

\date{\today}

\begin{abstract}
We present a derivation of Markovian master equation for the out of equilibrium quantum dot connected to two superconducting reservoirs, which are described by the Bogoliubov-de Gennes Hamiltonians and have the chemical potentials, the temperatures, and the complex order parameters as the relevant quantities. We consider a specific example in which the quantum dot is represented by the Anderson impurity model and study the transport properties, proximity effect and Andreev bound states in equilibrium as well as far from equilibrium setups. 
\end{abstract}


\maketitle

\section{Introduction}

Recent advancements of experimental techniques make it possible to fabricate nano-electronic devices where a quantum dot is connected to two superconducting electrodes.\cite{Winkelmann09}  Below the critical temperature, the electrons form a superconducting condensate (in other words, a single macroscopic quantum state). Therefore, in the case where the electrodes are superconducting, the quantum dot setup allows us to study the single electron  tunneling between two condensates held at different chemical potentials, temperatures, or being forced to have different order parameters (e.g. different phases of the anomalous electron density). The mixture of different physical phenomena, such as single electron tunneling, quantum phase transition, and macroscopic condensation, opens the possibilities to study the fundamental physics.\cite{yeyati11} 

The electron transport through a quantum dot   involves three different energy scales: the tunneling coupling between the dot  and the electrodes, the strength of  electronic correlations inside the dot, and the order parameter for the superconducting state in the electrodes. 
Most of the theoretical research that has been done so far has been employing Keldysh nonequilibrium Green's functions (NEGF) or scattering theory type approaches.\cite{PhysRevB.55.R6137,PhysRevB.73.214501,PhysRevB.77.104525}
NEGF and scattering theory  are able  to treat the tunneling coupling exactly, but they usually fully neglect correlations inside the dot or they rely on mean field or perturbation theory to treat them. Here we develop an approach which is based on the Markovian quantum master equation \cite{Lindblad76,GKS76}. The master equation approach to quantum transport works in the opposite regime -- it can treat the correlations inside the dot  very accurately (even exactly in the case of model systems) but the tunneling  is usually considered in the Born-Markov approximation.  Such an approach has been proved very useful for treating non-equlibrium transport problem in various quantum systems ~\cite{gurvitz96,PhysRevB.78.235424,PhysRevB.74.235309,PhysRevB.80.045309,PhysRevB.71.205304,PhysRevB.72.195330, Prosen2008,ProsenZunkovic2010,Prosen2011,dzhioev11b,dzhioev11a, KosovDzhioev2012}. It has been also applied to superconducting systems \cite{DubiVentra09}, where the proximity effect in one dimensional wires was studied. A Lindblad master equation with quadratic Lindblad operators was obtained in the mean field approximation by mapping the many body super-operator to a single particle form \cite{PershinDubiVentra08}. We note that a consistent treatment of the baths in the master equation approach poses many delicate issues \cite{Petruccione} (see e.g. also discussion in Ref.~\cite{ProsenZunkovic2010}). 

Here  we present a derivation of the master equation in the case when the electrodes are described by Bogoliubov-de Gennes Hamiltonians and then apply it to the non-equilibrium superconducting Anderson impurity model. We study in detail the transport properties of the model and the  proximity effect in quantum dot. Three different regimes are considered.  First, we focus on the generic case $2\epsilon+U\neq0$, where $\epsilon$ is the resonance level energy  and $U$ the interaction strength, where an exact, analytic expression for the steady state is found. In the particle-hole symmetric regime $2\epsilon+U=0$ we consider two cases, namely a dissipative one $\Delta<|\epsilon\pm\mu/2|$ and a non-dissipative one $\Delta>|\epsilon\pm\mu/2|$, where $\mu$ is the chemical potential bias and $\Delta$ the magnitude of the superconducting order parameter. In the dissipative case the phase difference dependent non-equilibrium particle current, the energy current, and the proximity effect are obtained. In the non-dissipative case, the Josephson current  originating from the Andreev bound states is discussed. The energies of the Andreev bound states and the corresponding particle current are obtained for arbitrary superconducting order parameter $\Delta$ and onsite energy level of the quantum dot $\epsilon$.

The paper is organized as follows.
In Sec.~\ref{master_equation}, we derive the master equation for a quantum system connected to superconducting baths, and then specialize on the specific derivation for  the out of equilibrium Anderson impurity model connected to the two superconducting leads.
In Sec.~\ref{solve_master}, we present the numerical and analytical solutions of the master equation for the model cases. 
Conclusions are given in Sec.~\ref{conclusions}. We use natural dimensionless units throughout the paper, in which $\hbar= k_B = |e| = 1$, where $-e$ is the electron charge.

\section{Markovian master equation for a quantum dot connected to superconducting baths }
\label{master_equation}

In the derivation of the Lindblad master equation one usually assumes that the interaction operators between the system and the bath are written in a Hermitian form.\cite{Petruccione} This is always possible and it usually simplifies the formal  derivation. Therefore, we begin with the outline of a general derivation of the Lindblad master equation and highlight the main differences from the usual textbook approach.\cite{Petruccione} The complete Hamiltonian 
is divided into three parts
\begin{equation}
H=H_{\rm S}+H_{\rm B}+H_{\rm I},
\end{equation}
where $H_{\rm B}$ denotes the bath Hamiltonian, $H_{\rm S}$ is  the system Hamiltonian, and $H_{\rm I}$  is the interaction between the system and the bath. The interaction can always be represented in the following separable form 
\begin{eqnarray}
\label{eq:H_i}
H_{\rm I}={\sum_{\alpha}} A_\alpha B_\alpha 
\end{eqnarray}
where the operators $A_\alpha$ (acting on the system) and $B_\alpha$ (acting on the bath) commute $[A_\alpha,B_{\alpha}]=0$. As noted before, we shall avoid the common assumption that $A_\alpha^\dag=A_\alpha$ and $B_\alpha^\dag=B_\alpha$,  since in our case the special form of the superconducting bath correlation functions induces two physically distinct contributions to the dissipator that are clearly separated only if we use the above form of the interaction (\ref{eq:H_i}). 
However, since $H_{\rm I}$ needs to be Hermitian, the set $\{A_\alpha B_\alpha\}$ has to include pairs of mutually Hermitian conjugate operators, i.e. for each $\alpha$ there exists $\alpha'$, such that
$A_{\alpha'} = A_\alpha^\dagger, B_{\alpha'}=B^\dagger_\alpha$.
The density matrix of the complete system satisfies the von Neumann equation.
We use the standard Born-Markov approximations, namely that the denisty matrix of the complete system can be written in a separable form $\rho(t)=\rho_{\rm S}\otimes\rho_{\rm B}$, where $\rho_{\rm S}$ denotes the density matrix of the system and $\rho_{\rm B}$ the density matrix of the bath, which is assumed to be in a Gibbs state. Therefore, we can simplify the von Neumann equation and trace out the bath degrees of freedom. Further, by performing an additional secular approximation we obtain the Lindblad master equation for the reduced density matrix of the system
\begin{eqnarray}
\label{eq:lindblad1}
\!\!\!\!\!\!\!\!\!\!\!\!\!\!\!\!\!\!\!\!\!\!\!\!\!\!\!\!\!\!\!\!\!\!\!\!\!\!\!\!\!\! \frac{{\rm d}\rho_{\rm S}(\tau)}{{\rm d}\tau} &=& -{\rm i}[H_{\rm LS},\rho_{\rm  S}(\tau)]+\hat{\mathcal{D}}\rho_{\rm  S}(\tau),\\ 
\label{eq:H_LS}
\!\!\!\!\!\!\!\!\!\!\!\!\!\!\!\!\!\!\!\!\!\!\!\!\!\!\!\!\!\!\!\!\!\!\!\!\!\!\!\!\!\! H_{\rm LS}&=&\sum_\omega\sum_{\alpha,\beta}S_{\alpha\beta}(\omega)\hat{\Pi}_{-\omega}(A_\alpha)\hat{\Pi}_{\omega}(A_\beta),\\ 
\label{eq:dissipator}
\!\!\!\!\!\!\!\!\!\!\!\!\!\!\!\!\!\!\!\!\!\!\!\!\!\!\!\!\!\!\!\!\!\!\!\!\!\!\!\!\!\! \hat{\mathcal{D}}\rho_{\rm  S}(\tau)&=&\sum_\omega\sum_{\alpha,\beta}\gamma_{\alpha\beta}(\omega)\left( 2\hat{\Pi}_\omega( A_\beta)\rho_{\rm  S}(\tau)\hat{\Pi}_{-\omega}(A_\alpha) - \left\{\hat{\Pi}_{-\omega}(A_\alpha)\hat{\Pi}_{\omega}(A_\beta),\rho_{S}(\tau)\right\}\right)\!\!,
\end{eqnarray}
where $[\bullet,\bullet]$ denotes the commutator and $\{\bullet,\bullet\}$ the anticommutator. The super-operators $\hat{\Pi}_\omega$ are projection super-operators on the eigenoperators of the  system Hamiltonian $H_{\rm S}$ and are defined as 
\begin{equation}
\hat{\Pi}_\omega(O_{\rm S})=\sum_{\epsilon'-\epsilon=\omega}|\epsilon\rangle\langle\epsilon|O_{\rm S}|\epsilon'\rangle \langle\epsilon'|,
\end{equation}
where $|\epsilon\rangle\langle\epsilon|$ are the projection operators on the possibly degenerate subspace of the system with the energy $\epsilon$ ($H_{\rm S} |\epsilon\rangle =\epsilon |\epsilon\rangle $) and $O_{\rm S}$ is an arbitrary operator acting on the system. Note, that the equation (\ref{eq:lindblad1}) can be brought to a standard Lindblad form since the matrix $\gamma_{\alpha\beta}$ is Hermitian and positive semi-definite.\cite{Petruccione}
The functions $S_{\alpha\beta}(\omega)$ and  $\gamma_{\alpha\beta}(\omega)$ are computed from the bath correlation function
\begin{eqnarray}
\label{eq:corr_def}
\!\!\!\!\!\!\!\!\!\!\!\!\!\!\!\!\!\!\!\!\!\!\!\!\!\!\!\!\!\!\!\!\!\!\!\!\!\!\!\! \Gamma_{\alpha\beta}(\omega)&=&\Gamma(B_{\alpha},B_{\beta}|\omega)= \int_0^\infty  {\rm d}s\, {\rm e}^{{\rm i}\omega s}\, {\rm tr}_{\rm B}\left( B_\alpha(\tau) B_\beta(\tau-s)\right)=\gamma_{\alpha\beta}(\omega)+{\rm i}S_{\alpha\beta}(\omega),\\
\!\!\!\!\!\!\!\!\!\!\!\!\!\!\!\!\!\!\!\!\!\!\!\!\!\!\!\!\!\!\!\!\!\!\!\!\!\!\!\! \gamma_{\alpha\beta}(\omega)&=&\gamma(B_{\alpha},B_{\beta}|\omega)=\frac{1}{2}\left(\Gamma(B_{\alpha},B_{\beta}|\omega)+\Gamma^*(B_{\beta}^\dag,B_{\alpha}^\dag|\omega)\right),\\
\!\!\!\!\!\!\!\!\!\!\!\!\!\!\!\!\!\!\!\!\!\!\!\!\!\!\!\!\!\!\!\!\!\!\!\!\!\!\!\! S_{\alpha\beta}(\omega)&=&S(B_{\alpha},B_{\beta}|\omega)=\frac{1}{2\rm i}\left(\Gamma(B_{\alpha},B_{\beta}|\omega)-\Gamma^*(B_{\beta}^\dag,B_{\alpha}^\dag|\omega)\right),
\end{eqnarray}
where ${\rm tr}_{\rm B}(\bullet)$ denotes a trace over the bath. 
\begin{figure} [!!h]
\centering{\includegraphics[width=0.405\textwidth]{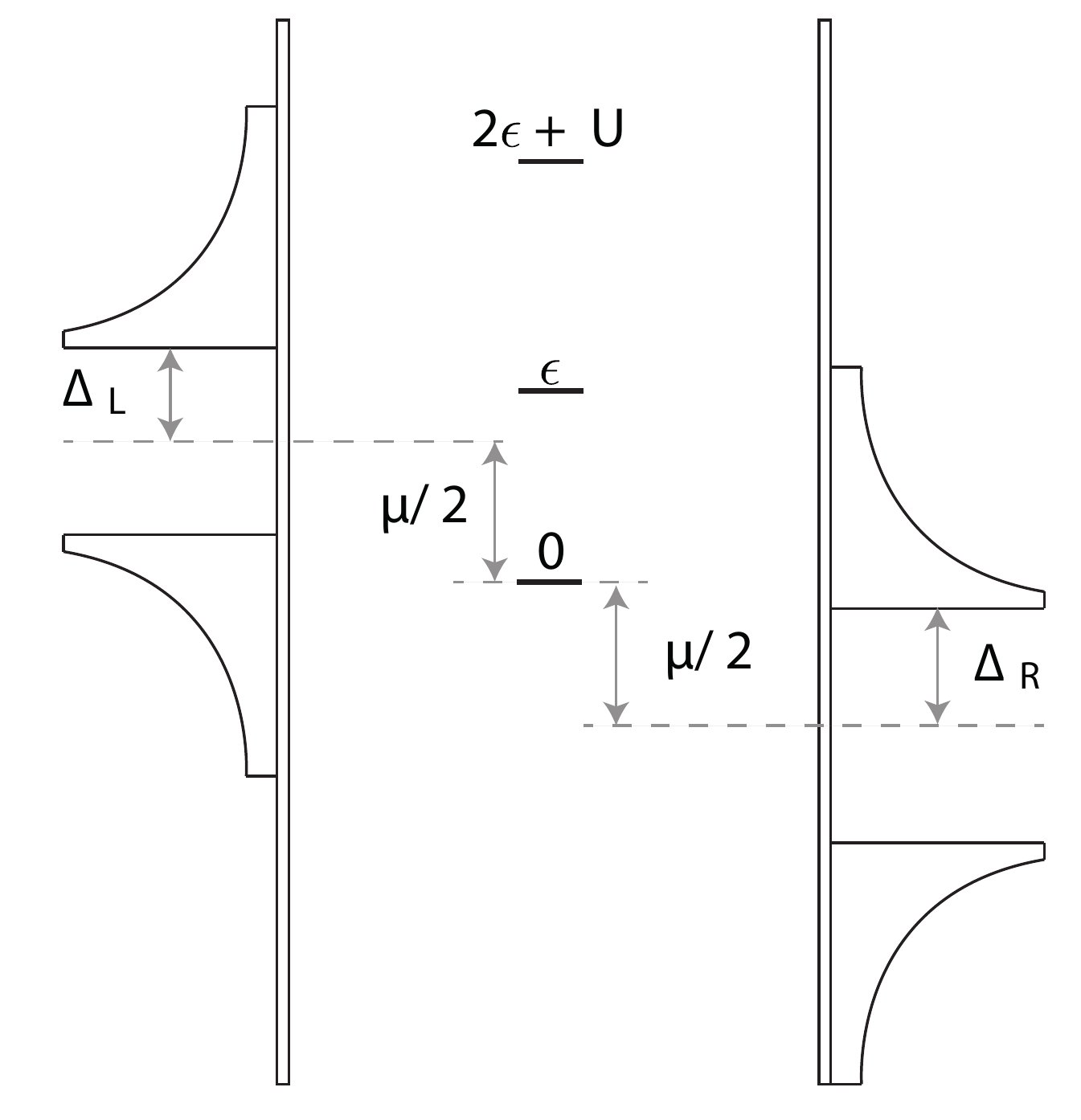}}
\caption{Schematic illustration  of the  out equilibrium superconducting Anderson impurity model: one spin-degenerate level with energy $\epsilon$ and local electronic repulsion $U$ is connected to two supeconducting semi-infinite leads described by the  Bogoliubov-de Gennes Hamiltonians. The chemical potential difference between left and right reservoir $\mu$ is included in the model through the bath correlation functions and describes the effect of a bias voltage, which is usually measured in experiment. In a similar manner the energy $\epsilon$ may be associated with the gate voltage. }
\label{fig:model}
\end{figure} 

Let us now apply the above consideration to  a specific model, shown in Fig. \ref{fig:model}. We consider a quantum dot connected to two uncorrelated one-dimensional superconconducting leads described by the  Bogoliubov-de Gennes Hamiltonian
\begin{equation}
\label{eq:setup}
\!\!\!\!\!\!\!\!\!\!\!\!\!\!\!\!\!\!\!\!\!\!\!\!\!\!\!\!\!\!H_{\rm B}=\sum_k\epsilon_k(b_{k,\uparrow}^\dag b_{k,\uparrow}+b_{-k,\downarrow}^\dag b_{-k,\downarrow})+\Delta({\rm e}^{{\rm i} \phi}b_{-k,\downarrow} b_{k,\uparrow}+{\rm e}^{-{\rm i} \phi}b_{k,\uparrow}^\dag b_{-k,\downarrow}^\dag).
\end{equation}
Here $b_{k,\sigma}^\dag /  b_{k,\sigma}$ are  creation/annihilation operators for an electron with spin $\sigma= \uparrow, \downarrow $ and  single-particle energy $\epsilon_k$,
$\Delta {\rm e}^{{\rm i\phi}} $ is the complex order parameter, which governs the supeconducting properties of the leads. 
The index $k$ runs over the modes of the left and right leads and $\Delta$ and phase $\phi$ may have different values for the left and right leads (but we do not wish to burden the notation with additional indices).
The quantum dot consists of one  spin-degenerate level with on-site energy $\epsilon$ and with local Coulomb interaction $U>0$:
\begin{equation}
  H_{\rm S}=\epsilon\sum_{\sigma}n_\sigma + Un_\uparrow n_\downarrow,
\end{equation}
where $n_\sigma=a^\dag_\sigma a_\sigma$ is the number operator  for electrons  with spin $\sigma$ in the quantum dot. Here $a^\dag_\sigma $ and $a_\sigma $ are creation and annihilation operators in the quantum dot, respectively. The interaction between the  quantum dot and the superconducting leads is taken to be in the standard tunneling form
\begin{equation}
\label{eq:interaction1a}
H_{\rm I} = \sum_{k\sigma}t_{k,\sigma}(b_{k,\sigma}^\dag a_\sigma+a_\sigma^\dag b_{k,\sigma}).
\end{equation}
Since we are dealing with fermions the creation and the annihilation operators in the bath and in the system anticommute $\{a_\sigma^{(\dag)},b_{k,\sigma'}^{(\dag)}\}=0$. To establishe the connection with the master equation derived above (\ref{eq:lindblad1}), where we assumed that system and bath operators in $H_{\rm I}$ commute with each other,  we perform a Jordan-Wigner rotation of fermonic creation and annihilation operators. Namely,  we identify the operators in the interaction part of the Hamiltonian $H_{\rm I}$ as $A_{\sigma} = a_{\sigma} P_{\rm B} $ acting on the system and the corresponding $B_{k,\sigma} = t_{k,\sigma} b_{k,\sigma}^{\dag} P_{\rm B}$ acting on the bath, where  $P_{\rm B} =\exp\left({\rm i}\pi\sum_{k,\sigma}b_{k,\sigma}^{\dag}b_{k,\sigma}\right)$ is the parity operator in the bath, which satisfies the following (anti)commutation relations
\begin{eqnarray}
[a_\sigma^{(\dag)},P_{\rm B}]&=&0,~~\{b^{(\dag)}_{k,\sigma},P_{\rm B}\}=0.
\end{eqnarray}
It is easy to verify that $A_{\sigma}$ and $ B_{k,\sigma_k}$ (for $k=1,2,\ldots$ and $\sigma,\sigma_k=\uparrow, \downarrow$) commute
\begin{eqnarray}
\!\!\!\!\!\!\!\!\!\!\!\!\!\!\!\!\!\!\!\!\!\!\!\!\!\!\!\!\!\!\!\!\!\!\!\!\! [A_{\sigma},B_{k,\sigma_k}]&=&[a_{\sigma} P_{\rm B}, t_{k,\sigma_k} b_{k,\sigma_k} P_{\rm B}]=t_{k,\sigma_k}(a_{\sigma} P_{\rm B} b_{k,\sigma_k} P_{\rm B}-b_{k,\sigma_k} P_{\rm B} a_{\sigma} P_{\rm B} )\\ \nonumber
\!\!\!\!\!\!\!\!\!\!\!\!\!\!\!\!\!\!\!\!\!\!\!\!\!\!\!\!\!\!\!\!\!\!\!\!\! &=&t_{k,\sigma_k}(a_{\sigma} P_{\rm B} b_{k,\sigma_k} P_{\rm B}-a_{\sigma} P_{\rm B} b_{k,\sigma_k} P_{\rm B})=0.
\end{eqnarray}
Hence, the interaction part of the Hamiltonian can be written as
\begin{eqnarray}
\label{eq:interaction1}
\!\!\!\!\!\!\!\!\!\!\!\!\!\!\!\!\!\!\!\!\!\!\!\!\!\!\!\!\!\!\!\!\!\!\!\!\!H_{\rm I}& =&  \sum_{k,\sigma}t_{k,\sigma}\left((b_{k,\sigma}^\dag P_{\rm B} )(P_{\rm B} a_\sigma)+(a_\sigma^\dag P_{\rm B})( P_{\rm B} b_{k,\sigma})\right)=\sum_{k,\sigma}( A_{\sigma} B_{k,\sigma} +A_{\sigma} ^\dag B_{k,\sigma} ^\dag).
\end{eqnarray}
Now we can calculate the correlation matrices (\ref{eq:corr_def}) for our model (see Appendix \ref{ap:calc_corr}). In order to obtain the dissipative part of the dynamics we have to find the projectors of the operators $A_{\sigma}$, $A_{\sigma}^\dag$ on the eigenoperators of the Hamiltonian as well. The eigenvectors (states) and the corresponding eigenvalues (energies) of the dot Hamiltonian are denoted as follows
\begin{eqnarray}
\begin{array}{|c||c|c|c|c|}
\hline
\mbox{States}&|0\rangle&|\!\!\uparrow\rangle=a_\uparrow^\dag|0\rangle&|\!\!\downarrow\rangle =a_\downarrow^\dag|0\rangle&|\!\!\uparrow\downarrow\rangle =a_\uparrow^\dag a^\dag_\downarrow|0\rangle\\ 
\hline
\mbox{Energies}&0&\epsilon&\epsilon&2\epsilon+U\\
\hline
\end{array}
\end{eqnarray}
Here the state $|0\rangle$ denotes the particle vacuum. Hence, the nonzero projections of the operators $A_{\sigma}$ and $A_{\sigma}^\dag$ on the eigenspace of the Hamiltonian are 
\begin{eqnarray}
\label{eq:proj_dot}
\hat{\Pi}_{\epsilon}(a_\sigma P_{\rm B})&=&|0\rangle\langle \sigma|,\quad\hat{\Pi}_{\epsilon+U}(a_\sigma P_{\rm B})=s_\sigma|\bar{\sigma}\rangle\langle\uparrow\downarrow\!\!|, \\ \nonumber
\hat{\Pi}_{-\epsilon}(P_{\rm B} a_\sigma^\dag)&=&|\sigma\rangle\langle 0|,\quad\hat{\Pi}_{-\epsilon-U}(P_{\rm B} a_\sigma^\dag)=s_\sigma|\!\!\uparrow\downarrow\rangle\langle\bar{\sigma}|,
\end{eqnarray}
were $s_\uparrow=1$, $s_\downarrow=-1$ and $\bar{\sigma}$ denotes the opposite spin of $\sigma$. Inserting the above projections (\ref{eq:proj_dot}) and the correlation functions calculated in the Appendix {\ref{ap:calc_corr}} into the master equation  (\ref{eq:lindblad1}) we obtain the dissipative part of the Liouvillean of the quantum dot connected to a superconducting reservoir
\begin{eqnarray}
\label{d1}
\hat{\mathcal{D}}^{(1)}(\rho)=\sum_{\sigma}&\Big(&\gamma^{(1)}(-\epsilon)\left( 2|\sigma\rangle\langle 0|\rho|0\rangle\langle\sigma|-\{|0\rangle\langle0|,\rho \} \right)\\ \nonumber
&+&\gamma^{(1)}(-\epsilon-U)\left( 2|\!\!\uparrow\downarrow\rangle\langle \bar{\sigma}|\rho|\bar{\sigma}\rangle\langle\uparrow\downarrow\!\!|-\{|\bar{\sigma}\rangle\langle\bar{\sigma}|,\rho \} \right)~~~~~~\\ \nonumber
&+&\gamma^{(1)}(\epsilon)\left( 2|0\rangle\langle \sigma|\rho|\sigma\rangle\langle0|-\{|\sigma\rangle\langle\sigma|,\rho \} \right)\\ \nonumber
&+&\gamma^{(1)}(\epsilon+U)\left( 2|\bar{\sigma}\rangle\langle \uparrow\downarrow\!\!|\rho|\!\!\uparrow\downarrow\rangle\langle\bar{\sigma}|-\{|\!\!\uparrow\downarrow\rangle\langle\uparrow\downarrow\!\!|,\rho \} \right)\Big)
\end{eqnarray}
and the Lamb shift term (see Ref.~\cite{Petruccione} for the definition) in the Hamiltonian
\begin{eqnarray}
\label{eq:H_LS 1}
\!\!\!\!\!\!\!\!\!\!\!\!\!\!H^{(1)}_{\rm LS}=\sum_{\sigma}&\Big(&S^{(1)}(-\epsilon)|0\rangle\langle 0|+S^{(1)}(-\epsilon-U)|\bar{\sigma}\rangle\langle\bar{\sigma} |\\ \nonumber
&+&S^{(1)}(\epsilon)|\sigma\rangle\langle \sigma|+S^{(1)}(\epsilon+U)|\!\!\uparrow\downarrow\rangle\langle \uparrow\downarrow\!\!| \Big).
\end{eqnarray}
In the particle-hole symmetric case ($2\epsilon+U=0$) we have an additional contribution to the Lamb shift and the dissipator. This is a consequence of two effects: (i) the non-vanishing superconducting correlation functions $\Gamma(P_{\rm B}b_{k,\uparrow}^\dag,P_{\rm B}b_{-k,\downarrow}^\dag|\omega)$ and $\Gamma(b_{k,\uparrow}P_{\rm B},b_{-k,\downarrow}P_{\rm B}|\omega)$, which signal a finite density of cooper-pairs and (ii) the twofold degeneracy of the energy zero in the dot, which ensures non-vanishing projections of the operators $A_{\sigma}$ and $A_{\sigma}^\dag$ on the eigenspaces of the Hamiltonian $H_{\rm S}$ with opposite energies. Therefore, products $\hat{\Pi}_{\epsilon}(a_\sigma P_{\rm B})\hat{\Pi}_{-\epsilon}(a_{\bar{\sigma}} P_{\rm B})$ and $\hat{\Pi}_{\epsilon}(P_{\rm B} a_\sigma^\dag)\hat{\Pi}_{-\epsilon}(P_{\rm B} a_{\bar{\sigma}}^\dag)$  appearing in the sums  (\ref{eq:H_LS}) and (\ref{eq:dissipator}) do not vanish as in the non-degenerate case ($2\epsilon+U\neq0$) and we obtain the following, additional contributions to the dissipator 
\begin{eqnarray}
\label{eq:dissipator 2}
\hat{\mathcal{D}}^{(2)}(\rho)=\sum_\sigma&\Big(& 2\gamma^{(2)}(\epsilon)|\!\!\uparrow\downarrow\rangle\langle\sigma|\rho|\sigma\rangle\langle0| \\ \nonumber
&+&\gamma^{(2)}(-\epsilon)\left( 2 |\bar{\sigma}\rangle\langle0|\rho|\!\!\uparrow\downarrow\rangle\langle\bar{\sigma}| -\{ |\!\!\uparrow\downarrow\rangle\langle0|,\rho \}\right)\\ \nonumber
&+&2\gamma^{(2)*}(\epsilon) |0\rangle\langle\bar{\sigma}|\rho |\bar{\sigma}\rangle\langle\uparrow\downarrow\!\!| \\ \nonumber
&+&\gamma^{(2)*}(-\epsilon)\left( 2|\sigma\rangle\langle\uparrow\downarrow\!\!|\rho|0\rangle\langle\sigma| -\{ |0\rangle\langle \uparrow\downarrow\!\!|,\rho \}\right) \Big)
\end{eqnarray}
and to the Lamb shift
\begin{eqnarray}
\label{eq:H_LS 2}
H^{(2)}_{\rm LS}=2\Big(S^{(2)}(-\epsilon)|\!\!\uparrow\downarrow \rangle\langle 0|+{S^{(2)}}^*(-\epsilon)|0\rangle\langle\uparrow\downarrow\!\!|\Big).
\end{eqnarray}
For the sake of simplicity the above expressions for dissipators (\ref{d1}), (\ref{eq:dissipator 2}) and the Lamb shifts (\ref{eq:H_LS 2}), (\ref{eq:H_LS 1}) are written for one bath only. The contribution of the second bath is identical and additive, so  the total dissipator and the Lamb shift become
\begin{eqnarray}
\hat{\mathcal{D}}&=&\hat{\mathcal{D}}_{\rm L}+\hat{\mathcal{D}}_{\rm R},\quad H_{\rm LS}=H_{\rm LS,L}+H_{\rm LS,R},\\ \nonumber
\hat{\mathcal{D}}_{\nu}&=&\hat{\mathcal{D}}_{\nu}^{(1)}+\hat{\mathcal{D}}_{\nu}^{(2)},\quad H_{\rm LS,\nu}=H^{(1)}_{LS,\rm \nu}+H^{(2)}_{LS,\rm \nu},
\end{eqnarray}
where $\nu={\rm L,R }$. Note that we neglect the broadening of the systems energy levels due to the coupling to the leads, i.e. the levels are infinitely narrow. The broadening can be included by hand setting for example $\eta=\kappa$ (see Appendix \ref{ap:calc_corr}) or by self-consistent treatment of the master equation as suggested in Ref.~\cite{EspositoGalperin10}.

\section{Solution of the master equation}
\label{solve_master}
In this section we shall find the steady state density matrix $\rho_{\rm NESS}$ of the master equation (\ref{eq:lindblad1}). First we consider the non-degenerate quantum dot, where we have only one contribution to the dissipator, namely (\ref{d1}), and the Liouville equation is simplified to a rate equation. An explicit analytic form of steady state is obtained. In the second subsection we consider the particle-hole symmetric case, where the steady state is calculated numerically. We find nontrivial non-equilibrium sub-gap dynamics due to the effect of the Lamb shift (\ref{eq:H_LS 2}). In both cases we discuss the non-equilibrium particle current and energy current defined as a change of the number of particles in the system and system's energy, respectively, due to the interaction with the left bath
\begin{eqnarray}
\label{eq:current_def_p}
J^n&=&\hat{\mathcal{D}}^{\rm H}_{\rm L}\big(n\big)+{\rm i}\big[H_{\rm LS,L},n\big],\quad n=\sum_{\sigma}a^\dag_\sigma a_\sigma, \\
\label{eq:current_def_e}
J^e&=&\hat{\mathcal{D}}^{\rm H}_{\rm L}\big(H_{\rm S}\big)+{\rm i}\big[H_{\rm LS,L},H_{\rm S}\big], 
\end{eqnarray}
where the superscript H denotes the Heisenberg representation of the superoperator $\hat{\mathcal{D}}^{\rm H}_{\rm L}$.  We also discuss the proximity effect, namely the cooper pair density in the quantum dot 
\begin{eqnarray}
\label{eq:prox}
\Delta_{\rm dot} {\rm e}^{\rm i \phi_{\rm dot}}=\langle a_\uparrow a_\downarrow \rangle.
\end{eqnarray}

\subsection{Non-dengerate quantum dot: $2 \epsilon + U \ne 0$}
\begin{figure} [!!h]
\centering{\includegraphics[width=0.50\textwidth]{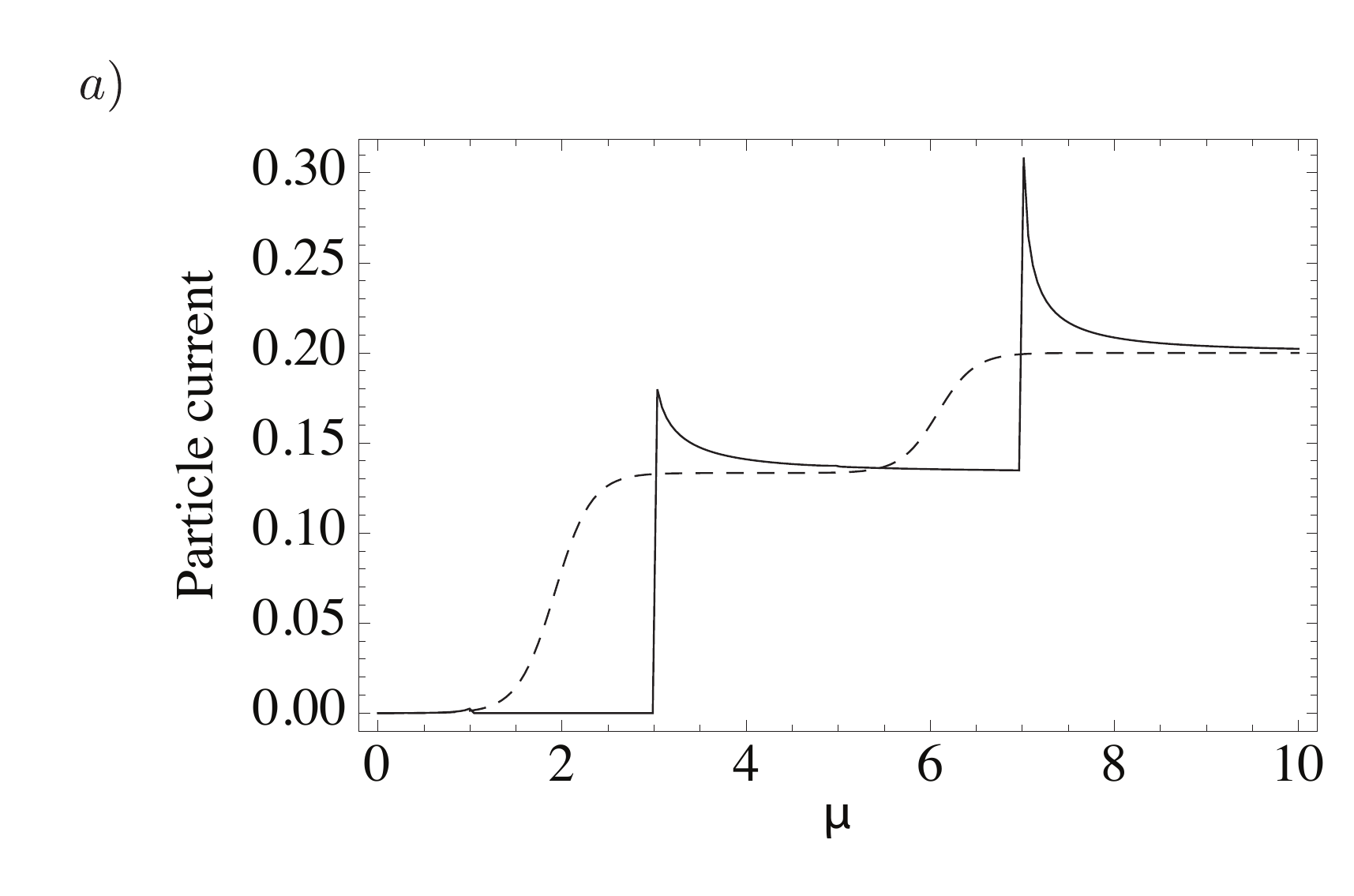}~~~~~~~~\includegraphics[width=0.43\textwidth]{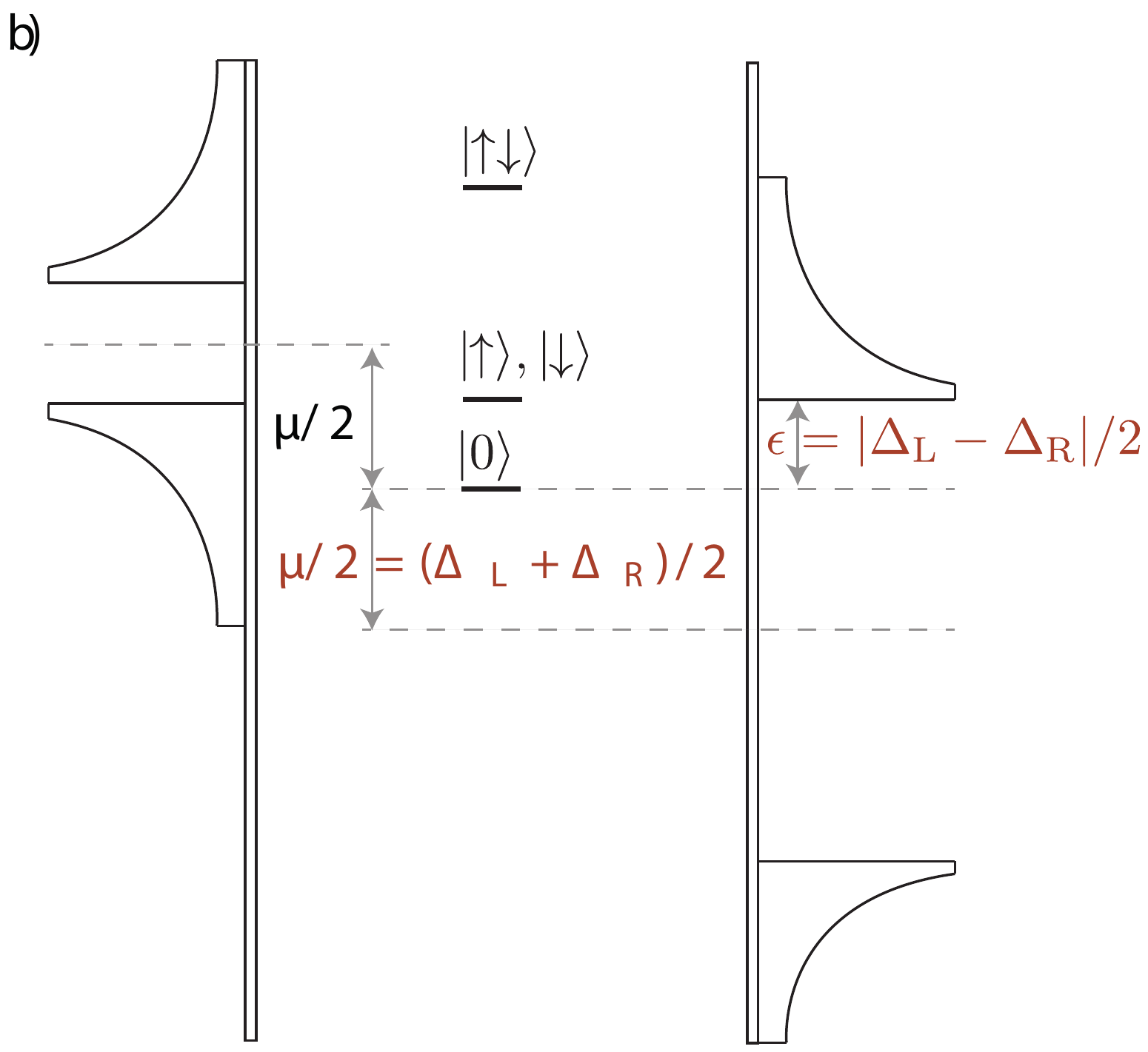}}
\caption{a) Current-voltage characteristics of the Anderson impurity. Dashed line corresponds to the  normal leads ($\Delta_{\rm L,R}=0$) and the full line corresponds to the superconducting leads ($\Delta_{\rm L,R}=0.5$). For superconducting leads the shift of the step by $2\Delta$ and the negative differential conductance are a consequence of the SDOS. Other model parameters: $\epsilon=1, U=2, T_{\rm L,R}=0.1, \Gamma=0.1\,.$ b) Schematic representation of the configuration with the maximal current. The chemical potential is  $\mu=\Delta_{\rm L}+\Delta_{\rm R}$, therefore the top of right  superconducting gap aligns with the bottom of the left superconducting gap; they align at the energy $\epsilon=|\Delta_{\rm L}-\Delta_{\rm R}|/2$.}
\label{fig:mu_jn}
\end{figure}

\begin{figure} [!!h]
\centering{ \includegraphics[width=0.256\textwidth]{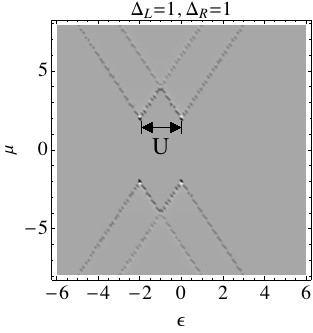} \includegraphics[width=0.24\textwidth]{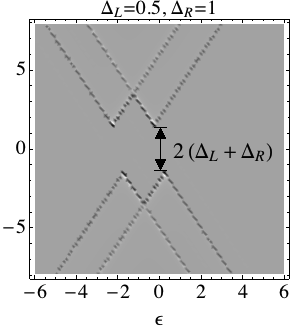}
\includegraphics[width=0.24\textwidth]{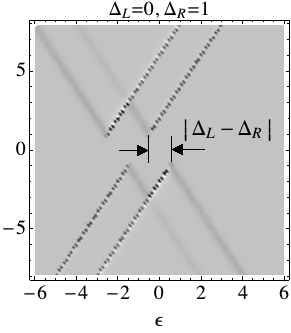} \includegraphics[width=0.24\textwidth]{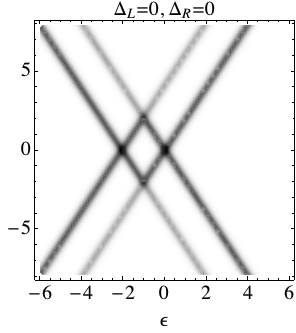}}
\caption{Density plot of differential conductance ($G=\partial\langle J^n\rangle/ \partial \mu$) as a function of the level position $\epsilon$ and the bias voltage (chemical potential difference) $\mu$, for different values of the order parameters $\Delta_{\rm L,R}$ (indicated on top of each panel). The characteristic distances denoted in the first three panels are calculated from the equation (\ref{eq:G_peak}); see main text and panel b) of Fig. \ref{fig:mu_jn}.  Dark/light gray represent high/low values of differential conductance $G$ with, in each plot, suitably adjusted relative scale. Other model parameters: $U=2, T_{\rm L,R}=0.1, \Gamma=0.1$.}
\label{fig:eps_mu_G}
\end{figure}

As already explained, in the non-degenerate case we need to take into account only the first part of the dissipator  ($\hat{\mathcal{D}}^{(1)}$), Eq. (\ref{d1}). The subindex $\nu={\rm L,R}$ in the correlation functions $\gamma^{(j)}_\nu$ and $S^{(j)}_\nu$ denotes different baths. In this case the steady state can be found analytically by writing the Liouvillean in the matrix form and noting that the coherences decouple from the rates, which results in a simple rate equation the solution of which is
\begin{eqnarray}
\label{rhoness}
\!\!\!\!\!\!\!\!\!\!\!\!\rho_{\rm NESS}&=&(\rho_0|0\rangle\langle0|+\rho_1(|\!\!\uparrow\rangle\langle\uparrow\!\!|+|\!\!\downarrow\rangle\langle\downarrow\!\!|)+\rho_2|\!\!\uparrow\downarrow\rangle\langle\uparrow\downarrow\!\!|)/(\rho_0+2\rho_1+\rho_2),\\ \nonumber
\!\!\!\!\!\!\!\!\!\!\!\!\rho_0&=&\left(\gamma_{\rm L}^{(1)}(\epsilon )+\gamma_{\rm R}^{(1)}(\epsilon )\right) \left(\gamma_{\rm L}^{(1)}(U+\epsilon )+\gamma_{\rm R}^{(1)}(U+\epsilon )\right),\\ \nonumber
\!\!\!\!\!\!\!\!\!\!\!\!\rho_1&=& \left(\gamma_{\rm L}^{(1)}(-\epsilon )+\gamma_{\rm R}^{(1)}(-\epsilon )\right)\left(\gamma_{\rm L}^{(1)}(U+\epsilon )+\gamma_{\rm R}^{(1)}(U+\epsilon )\right),\\ \nonumber
\!\!\!\!\!\!\!\!\!\!\!\!\rho_2&=& \left(\gamma_{\rm L}^{(1)}(-\epsilon )+\gamma_{\rm R}^{(1)}(-\epsilon )\right) \left(\gamma_{\rm L}^{(1)}(-U-\epsilon )+\gamma_{\rm R}^{(1)}(-U-\epsilon )\right).
\end{eqnarray}
Interesting observables in the non-degenerate case are the particle current (\ref{eq:current_def_p}) and the energy current (\ref{eq:current_def_e}), which simplify to
\begin{eqnarray}
\label{eq:currents1}
\!\!\!\!\!\!\!\!\!\!\!\!\!\!\!\!\!\! J^n=\sum_\sigma&\left(-4\gamma_{\rm L}^{(1)}(-\epsilon)|0\rangle\langle0|+\left(-2\gamma_{\rm L}^{(1)}(-\epsilon-U)+2\gamma_{\rm L}^{(1)}(\epsilon)\right)|\sigma\rangle\langle\sigma| \right. \\ \nonumber
&\left.+4\gamma_{\rm L}^{(1)}(\epsilon+U)|\!\!\uparrow\downarrow\rangle\langle\uparrow\downarrow\!\!| \right),
\end{eqnarray}
and
\begin{eqnarray}
\label{eq:currents2}
\!\!\!\!\!\!\!\!\!\!\!\!\!\!\!\!\!\!J^e=\sum_\sigma&&\left(-4\epsilon\gamma_{\rm L}^{(1)}(-\epsilon)|0\rangle\langle0|+\left(-2U\gamma_{\rm L}^{(1)}(-\epsilon-U)+2\epsilon\gamma_{\rm L}^{(1)}(\epsilon)\right)|\sigma\rangle\langle\sigma|\right.\\ \nonumber
&&\left.+4(U+\epsilon)\gamma_{\rm L}^{(1)}(\epsilon+U)|\!\!\uparrow\downarrow\rangle\langle\uparrow\downarrow\!\!| \right),
\end{eqnarray}
respectively, where the subscript (L) denotes the left bath.  By using the equations (\ref{rhoness}), (\ref{eq:currents1}), and (\ref{eq:currents2}) we can easily calculate the expectation values of the particle and energy currents in the steady state. 
The chemical potential is included by replacing $\epsilon\rightarrow\epsilon\pm\mu/2$ in the left (+) and the right (-) bath correlation functions.  The qualitative behavior of the currents (and the differential conductance) can entirely be explained by the electronic density of states in the superconducting leads (superconducting density of states - SDOS),
\begin{eqnarray}
\rho_L(\omega)=\Theta(|\omega|-\Delta)|\omega|/\sqrt{\omega^2-\Delta^2},
\end{eqnarray} 
where $\Theta(\omega)$ is the Heaviside step function (shown in Fig. \ref{fig:model}). The coupling strength to the baths $\gamma^{(1)}_{\rm L,R}$ is proportional to the SDOS, as shown in the Appendix \ref{ap:calc_corr} . Hence, the main features of SDOS, namely a gap $2\Delta$ where the electronic density of states is zero and a divergence at the border of the gap are reflected in the current-voltage characteristics (see Fig. \ref{fig:mu_jn}) and the differential conductance $G=\partial \langle J^n\rangle/ \partial \mu$  map  (see Fig. \ref{fig:eps_mu_G}) . Far from the gap, i.e.  when $\Delta_{\rm L,R}\ll |\epsilon\pm\mu/2|$ and $\Delta_{\rm L,R}\ll |\epsilon\pm\mu/2+U|$, the current approaches the value calculated for the normal leads. As we approach the superconducting gap (by changing the chemical potential $\mu$ or the  onsite energy $\epsilon$) we observe a peak in the differential conductance when one set of the following conditions is satisfied
\begin{eqnarray}
\!\!\!\!\!\!\!\!\!\!\!\!\!\!\!\!\!\!\!\!\!\!\!\!\!\!\!\!\!\!\!\!\omega=\Delta_{\rm L}-\mu/2\mbox{~and~}\omega\leq-\Delta_{\rm R}+\mu/2 \mbox{~or~}\omega\geq\Delta_{\rm L}-\mu/2\mbox{~and~}\omega=-\Delta_{\rm R}+\mu/2,
\label{eq:G_peak}
\end{eqnarray}
where $\omega=\epsilon$ or $\omega=\epsilon+U$ is the transition frequency between the subsequent levels of the dot. The conditions (\ref{eq:G_peak}) are valid if $\mu>0$, but for $\mu<0$ the roles of the baths are exchanged ($\Delta_{\rm L}\leftrightarrow\Delta_{\rm R}$). After the peak we observe negative differential conductance as a consequence of the decreasing density of states, which is clearly shown in Fig. \ref{fig:mu_jn}. The characteristic distances appearing in the differential conductance map in Fig. \ref{fig:eps_mu_G} can easily be calculated from the first and the last condition in (\ref{eq:G_peak}), namely the interaction energy $U$,  the sum of the superconducting order parameters in left and right lead  $\Delta_{\rm L}+\Delta_{\rm R}$, and the difference of the superconducting order parameters in the leads $|\Delta_{\rm L}-\Delta_{\rm R}|$. The interaction energy determines the difference between the two possible transition energies $\omega$ for one particle transfer and thus also the relative shift of the diamond structures on the  $\epsilon$ (gate voltage) axis in Fig. \ref{fig:eps_mu_G}. The distance $\Delta_{\rm L}+\Delta_{\rm R}$ determines the  bias voltage (or chemical potential) that has to be applied to the leads in order to get the maximal particle current, namely to align the top of one superconducting gap with the bottom of the other; they align at the energy $|\Delta_{\rm L}-\Delta_{\rm R}|/2$, which is shown in panel b) of Fig. \ref{fig:mu_jn}.

Note that, the cooper pair density in the quantum dot (the proximity effect) vanishes since there are no coherences in the steady state of the non-degenerate dot. If the level is placed inside of the superconducting gap of both superconductors the dot is not coupled to the environment and we obtain a trivial unitary evolution. In the next subsection we shall show that in the particle hole symmetric case the evolution of the level inside of the gaps is changed due to an additional non-vanishing Lamb shift term.

\subsection{Particle-hole symmetric case $2 \epsilon + U = 0$}
\begin{figure} [!!h]
\centering{\includegraphics[width=0.510\textwidth]{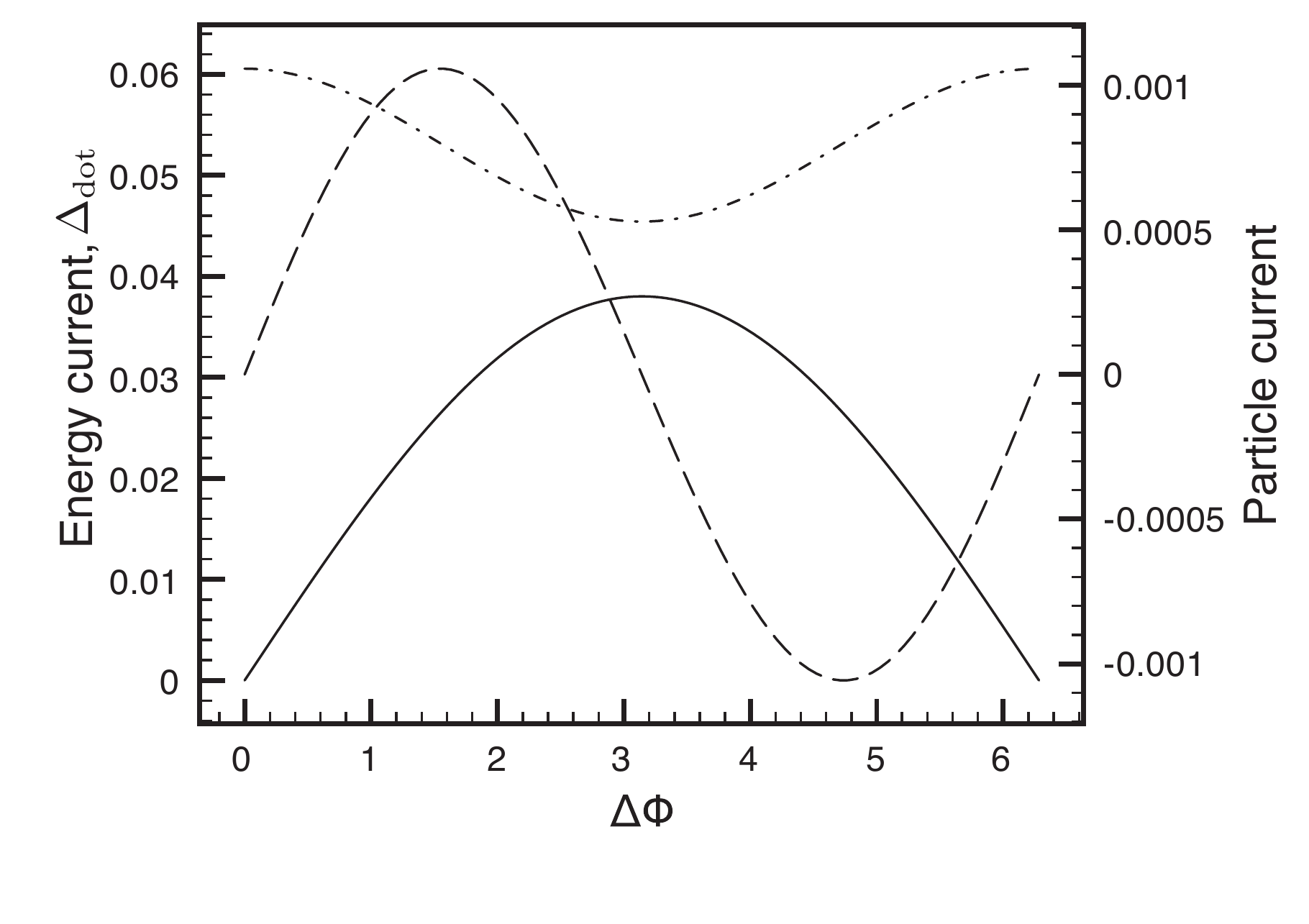}~~\includegraphics[width=0.51\textwidth]{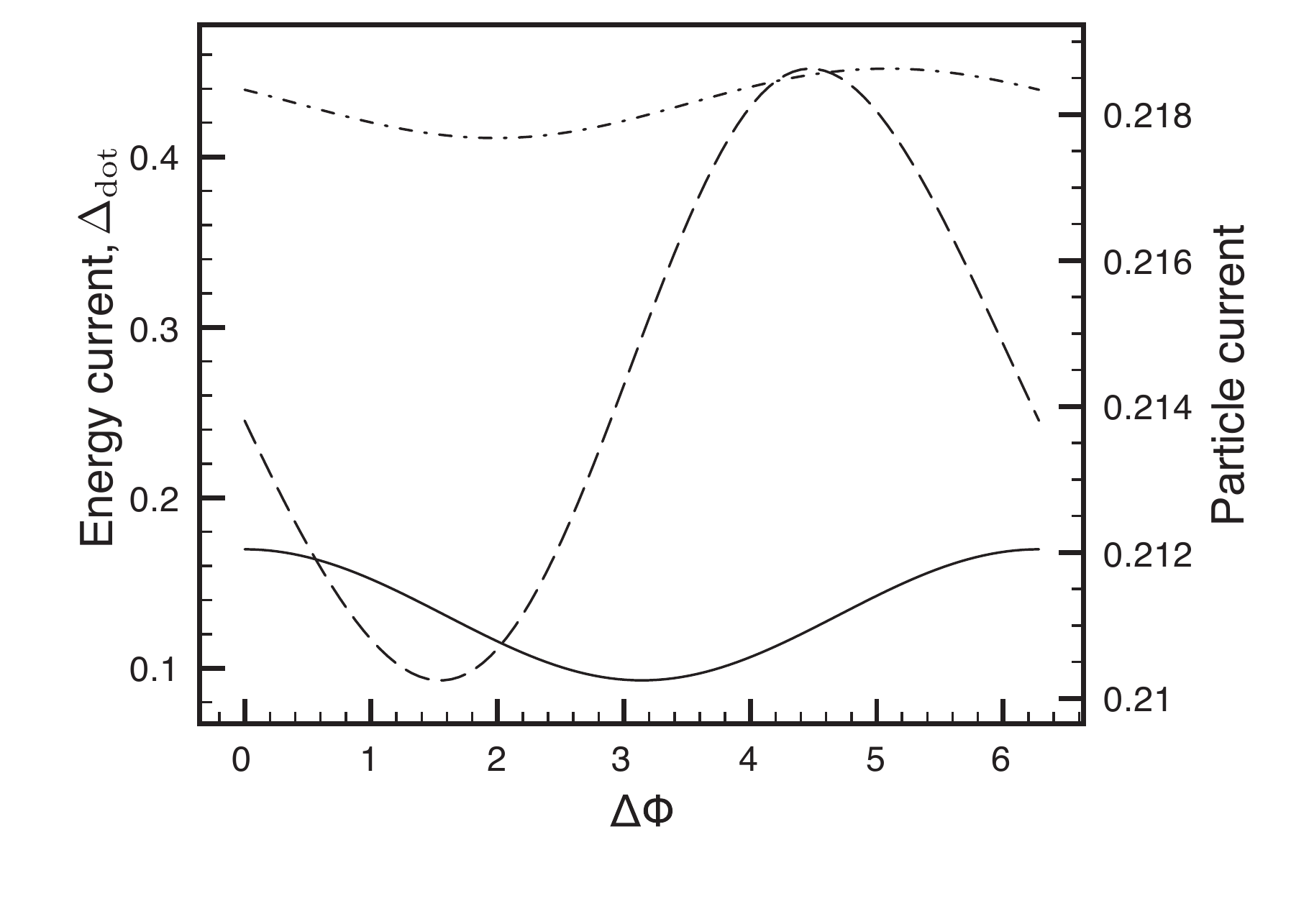}}
\caption{Out of equilibrium phase difference $\Delta\phi=\phi_{\rm L}-\phi_{\rm R}$ dependent particle current  $\langle J^n\rangle$ (dashed line), energy current $\langle J^e\rangle$ (dash dotted line) and proximity effect ($\Delta_{\rm dot}=|\langle a_\uparrow a_\downarrow \rangle|$, full line). The left panel is calculated for a temperature bias ($ \mu=0, T_{\rm L}=0.2, T_{\rm R}=1$) and the right panel shows the results obtained for a nonzero bias voltage (or chemical potential; $ \mu=4, T_{\rm L,R}=0.2$). Note, that if  $T_{\rm L}=T_{\rm R}$ and $\mu=0$ the particle current, energy current, and the Cooper pair density in the quantum dot vanish, therefore the obtained phase dependence is a purely non-equilibrium effect. Other model parameters: $2\epsilon=-U=-2, \Gamma=0.1, \Delta_{\rm L,R}=0.5$.}
\label{fig:neq_current}
\end{figure}

\begin{figure} [!!h]
\includegraphics[width=0.45\textwidth]{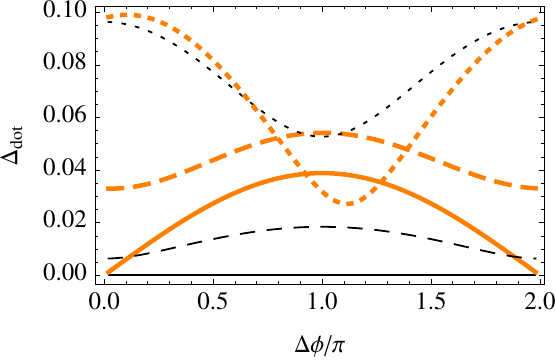} ~~~~\includegraphics[width=0.44\textwidth]{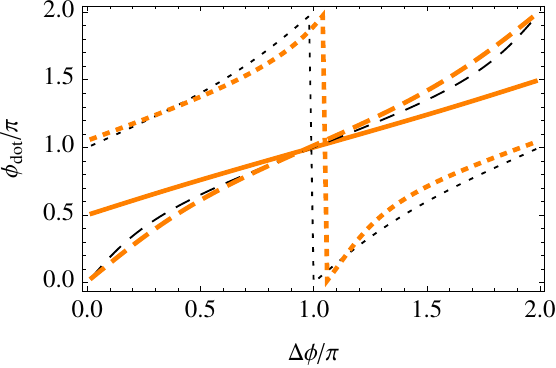}
\caption{ The proximity effect in quantum dot $\Delta_{\rm dot}{\rm e}^{\rm i \phi_{\rm dot}}=\langle a_\uparrow a_\downarrow\rangle$ for various regimes. The color and thickness distinguishes between the temperatures of the right bath (thin, black) $T_{\rm R}=1$, (thick, orange) $T_{\rm R}=0.1$ and the dashing represents the chemical potential (full line) $\mu=0$, (dashed lines) $\mu=0.5$, (dotted lines) $\mu=4$.  The curves calculated for large chemical potential bias are shifted by $\pi$ relative to the curves calculated for small chemical potential bias. An additional small phase shift occurs if a chemical potential bias and a temperature bias are present.  Other model parameters:  $\epsilon=-1, \Gamma=0.1, T_{\rm L}=1,  \Delta_{\rm L,R}=0.5$.}
\label{fig:proximity}
\end{figure}

In the symmetric case ($2\epsilon+U=0$)  an additional nontrivial term appears in the Lindblad master equation, which is not present for non-superconducting leads and describes Cooper pair tunneling between the bath and the dot. Not surprising, the dependence of the time evolution on the phase of superconducting order parameters in the leads is introduced through this second part of the dissipator (\ref{eq:dissipator 2}).  An analytic solution in this case is cumbersome, since the populations are now coupled to the coherences through the extra dissipator, therefore we rely  on numerical calculations to find the exact steady state. 
As in the non-degenerate case we study the particle current, energy current and the proximity effect.
It is interesting, that in equilibrium we have no particle current although the superconducting parameters in the left and the right lead have a different phase. The energy current and the Cooper pair density in the dot are zero as well. However, in out of equilibrium steady state the currents depend on the difference of the superconducting phases in the leads $\Delta\phi=\phi_{\rm L}-\phi_{\rm R}$  (see Fig. \ref{fig:neq_current}). Moreover, a finite non-equilibrium proximity effect (\ref{eq:prox}) is obtained; also shown in Fig. \ref{fig:neq_current}. Although we are unable to derive exact analytic expressions for the observed quantities we find some interesting effects when passing form the small bias regime $\mu<2(|\epsilon|+\Delta)$, where $\Delta_{\rm L}=\Delta_{\rm R}=\Delta$, to the large bias voltage regime $\mu>2(|\epsilon|+\Delta)$. 
If we change the chemical potential difference from small to large values we observe a phase flip, which is  $\pi-$shift in the phase dependence of the order parameter in the quantum dot. For zero bias voltage the phase profile is linear and it is antisymmetric under the reflection around the point $(\Delta\phi , \phi_{\rm dot})=(\pi,\pi)$. This symmetry in the phase dependence $\phi_{\rm dot}(\Delta\phi)$ is present whenever the temperature difference or the chemical potential difference between the baths is zero. This is illustrated in Fig. {\ref{fig:proximity}}, where we show the dependence of the phase $\phi_{\rm dot}$ and the amplitude $\Delta_{\rm dot}$ of the complex order parameter in the quantum dot  on the superconducting phase difference $\Delta\phi$. 

\begin{figure} [!!h]
\centering{\includegraphics[width=0.43\textwidth]{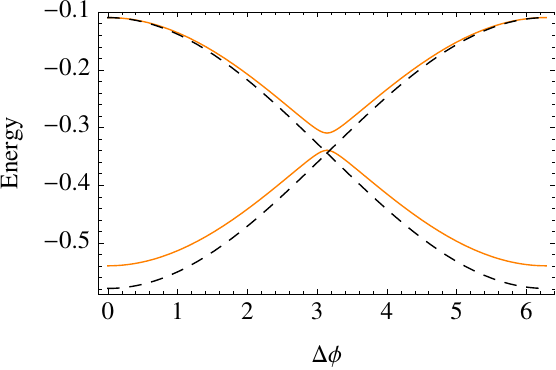} ~~~~\includegraphics[width=0.46\textwidth]{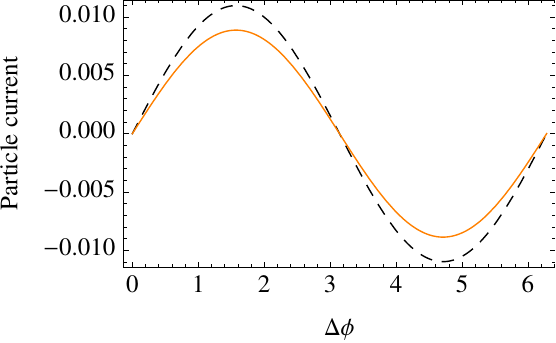}}
\caption{ We plot current carrying energy levels of the Lamb shift modified Hamiltonian $H_{\rm sym}$ (left) and the corresponding particle current  (right) calculated in the Gibbs state of $H_{\rm sym}$. Black dashed lines show the equilibrium result, namely the analytic result of equations (\ref{eq:andreev_states}) and (\ref{eq:jn_abs}), calculated for $\Delta_{\rm L,R}=2, \mu=0$. Out of equilibrium (orange lines) crossing of Andreev bound state energies is avoided (parameters: $\Delta_{\rm L}=2, \Delta_{\rm R}=5, \mu=1$). Note, that in case of nonzero chemical potential difference or bias voltage (orange lines) the particle current oscillates around zero. Other model parameters:  $\epsilon=-1/2, \Gamma=0.1, T_{\rm L,R}=1$.}
\label{fig:josephson_current}
\end{figure}

Next we consider the non-dissipative case, namely a  narrow  dot level (i.e. when the life time of the electron on the level is much larger than $1/2 \Delta $) is placed inside the superconducting gap below the chemical potential of the baths. Surprisingly, we observe a finite non-dissipative particle current, which comes from the distortion of the Hamiltonian due to the coupling to the baths i.e. from the Lamb shift. The modified Hamiltonian in the described case is 
\begin{eqnarray}
\!\!\!\!\!\!\!\!\!\!\!\!\!\!\!\!H_{\rm sym}&=&\left(
\begin{array}{cccc}
2 S^{(1)}(-\epsilon ) & 0 & 0 & 2{S^{(2)}}^*(-\epsilon) \\
 0 & \epsilon +2S^{(1)}(\epsilon ) & 0 & 0 \\
 0 & 0 & \epsilon +2S^{(1)}(\epsilon ) & 0 \\
 2S^{(2)}(-\epsilon) & 0 & 0 &2 S^{(1)}(-\epsilon )
\end{array}
\right),\\ \nonumber
\!\!\!\!\!\!\!\!\!\!\!\!\!\!\!\!S^{(j)}(\omega)&=&S^{(j)}_{\rm L}(\omega)+S^{(j)}_{\rm R}(\omega),\quad j=1,2.
\end{eqnarray}
The Hamiltonian in the degenerate subspace spanned by the states $|0\rangle$ and $|\uparrow\downarrow\rangle$ is perturbed by the interaction with the environment. This lifts the degeneracy and obtained energy eigenstates
\begin{eqnarray}
\label{eq:andreev_states}
\!\!\!\!\!\!\!\!\!\!\!\!\!\!\!\!\!\!\!\!\!\!\!\!\!\!\!\!\!\!\!\!\!\!\!\!\!\!\! |\Psi_\pm\rangle=\frac{1}{\sqrt{2}}\left(\pm e^{-i \Delta\phi /2 }|\!\!\uparrow\downarrow\rangle+|0\rangle\right),\quad E_\pm=2S^{(1)}(-\epsilon)\pm 4|S^{(2)}(-\epsilon)| \cos \left(\frac{\Delta\phi}{2}\right)
\end{eqnarray}
are no longer eigenstates of the particle number operator. The particle current in this case simplifies to
\begin{eqnarray}
J^n=4{\rm i}\left( -S^{(2)}_{\rm L}(-\epsilon)|\!\!\uparrow\downarrow\rangle\langle0| +{S^{(2)}_{\rm L}}^*(-\epsilon)|0\rangle\langle\uparrow\downarrow\!\!|\right).
\end{eqnarray}
and the expectation value of the particle current in the states $|\Psi_\pm\rangle$ is
\begin{eqnarray}
\langle\Psi_\pm|J^{\rm n}|\Psi_\pm\rangle=\mp 2  |S^{(2)}(-\epsilon)| \sin \left(\frac{\Delta\phi}{2}\right).
\label{eq:jn_abs}
\end{eqnarray}
In the above equations (\ref{eq:andreev_states}, \ref{eq:jn_abs}) we assume that $\Delta_{\rm L}=\Delta_{\rm R}=\Delta$ and $|\epsilon|=U/2<\Delta$. The current carying states can be interpreted as the Andreev bound states (see Ref. \cite{yeyati11}). Further, we assume that the dot is in the Gibbs state $\rho_G$ of the modified system Hamiltonian $H_{\rm sym} = H_{\rm S} + H_{\rm LS}$  and calculate the particle current (see Fig.~\ref{fig:josephson_current}). Interestingly, the particle current  oscillates around zero also in the case where we have a non-zero chemical potential difference.

\section{Conclusions}
\label{conclusions}

We derived a master equation for the electron transport through the quantum dot connected to two superconducting leads. In our derivations the Born-Markov and the rotating-wave approximations were used, which reduce the master equation to the standard Lidblad form with nontrivial dissipators and the Lamb shift terms originating from the superconducting baths. Then, the master equation was explicitly solved for out of equilibrium Anderson impurity model and the exact steady state density matrix  in the generic regime $2\epsilon+U \neq 0$ was found. In the particle-hole symmetric regime $2\epsilon+U=0$ a phase dependent dissipator was obtained. Surprisingly, the equilibrium solution in this case does not exhibit a superconducting  phase difference dependent particle current, whereas in the out of equilibrium steady state the particle current, the energy current, and the Cooper pair density of states in the quantum dot depend on the difference of the superconducting phases in the baths $\Delta\phi$. In the sub-gap case, Andreev bound states\cite{andreev64} are found as eigenstates of the Lamb shift perturbed Hamiltonian. Their energies and the corresponding non-dissipative particle current were obtained also in the non-equilibrium situation. The master equation derived in this article can be extended to treat larger systems, e.g. the double quantum dot, molecules, and one dimensional wires.

This work has been supported by the Francqui Foundation, Programme d'Actions de Recherche Concert\'ee de la Communaut\'e francaise (Belgium) under project ``Theoretical and experimental approaches to surface reactions",  the grants P1-0044 and J1-2208 of Slovenian Research Agency (ARRS), and a scholarship from the Slovene human resources developement and scholarship found.

\section*{References}
\bibliography{bcs}

\newpage

\appendix
\section{Calculation of correlation function  for supeconducting bath}
\label{ap:calc_corr}

In the appendix we shall obtain the correlation function in the bath with the Bogolyubov-de Gennes Hamiltonian, which can be diagonalized by the Bogolyubov transformation
\begin{eqnarray}
\!\!\!\!\!\!\!\!\!\!\!\!\!\!\!\!\!\!\!\!\!\!\!\!\!\!\!\!\!\!\!\! b_{k,\uparrow}&=&-u(\epsilon_k)d_{k,\uparrow}+v(\epsilon_k)d_{k,\downarrow}^\dag,\quad b_{-k,\downarrow}=u(\epsilon_k)d_{k,\downarrow}+v(\epsilon_k)d_{k,\uparrow}^\dag\\ \nonumber
\!\!\!\!\!\!\!\!\!\!\!\!\!\!\!\!\!\!\!\!\!\!\!\!\!\!\!\!\!\!\!\! \quad u(\epsilon)&=&{\rm e}^{-{\rm i}\phi}\sqrt{\frac{1}{2}+\frac{\epsilon}{2\omega(\epsilon)}},\quad v(\epsilon)=\sqrt{\frac{1}{2}-\frac{\epsilon}{2\omega(\epsilon)}}\\ \nonumber
\!\!\!\!\!\!\!\!\!\!\!\!\!\!\!\!\!\!\!\!\!\!\!\!\!\!\!\!\!\!\!\! \omega(\epsilon_k)&=&\sqrt{\epsilon_k^2+\Delta^2},\quad \tan(2 \phi_k)=-\frac{\Delta}{\epsilon_k},\\ \nonumber
\!\!\!\!\!\!\!\!\!\!\!\!\!\!\!\!\!\!\!\!\!\!\!\!\!\!\!\!\!\!\!\! H_{\rm B}&=&\sum_k\sum_{\sigma}\omega(\epsilon_k)d_{k,\sigma}^\dag d_{k,\sigma}.
\end{eqnarray}
We assume that the bath is in the equilibrium at temperature $T$
\begin{eqnarray}
\label{eq:ocupation}
\langle d_{k,\sigma}^\dag d_{k',\sigma'}\rangle_{\rm B}=n(\epsilon_k)\delta_{k,k'}\delta_{\sigma,\sigma'},\quad n(\epsilon_k)=\frac{1}{1+{\rm e}^{\omega(\epsilon_k)/T}}.
\end{eqnarray}
Hence, the only nonzero correlation functions are
\begin{eqnarray}
\label{eq:ap_gamma}
\!\!\!\!\!\!\!\!\!\!\!\!\!\!\!\!\!\!\!\!\!\!\!\!\!\!\!\!\!\!\!\!\Gamma(P_{\rm B}b_{k,\sigma}^\dag,b_{k,\sigma}P_{\rm B}|\omega)&=&{\rm i}|u(\epsilon_k)|^2\frac{n(\epsilon_k)}{\omega+\omega(\epsilon_k)+{\rm i}\eta}+{\rm i}|v(\epsilon_k)|^2\frac{1-n(\epsilon_k)}{\omega-\omega(\epsilon_k)+{\rm i}\eta},\\ \nonumber
\!\!\!\!\!\!\!\!\!\!\!\!\!\!\!\!\!\!\!\!\!\!\!\!\!\!\!\!\!\!\!\!\Gamma(b_{k,\sigma}P_{\rm B},P_{\rm B}b_{k,\sigma}^\dag|\omega)&=&{\rm i}|v(\epsilon_k)|^2\frac{n(\epsilon_k)}{\omega+\omega(\epsilon_k)+{\rm i}\eta}+{\rm i}|u(\epsilon_k)|^2\frac{1-n(\epsilon_k)}{\omega-\omega(\epsilon_k)+{\rm i}\eta},\\ \nonumber
\!\!\!\!\!\!\!\!\!\!\!\!\!\!\!\!\!\!\!\!\!\!\!\!\!\!\!\!\!\!\!\!\Gamma(b_{k,\uparrow}P_{\rm B},b_{-k,\downarrow}P_{\rm B}|\omega)&=&-{\rm i}u(\epsilon_k)v(\epsilon_k)\left(\frac{n(\epsilon_k)}{\omega+\omega(\epsilon_k)+{\rm i}\eta}-\frac{1-n(\epsilon_k)}{\omega-\omega(\epsilon_k)+{\rm i}\eta}\right),\\ \nonumber
\!\!\!\!\!\!\!\!\!\!\!\!\!\!\!\!\!\!\!\!\!\!\!\!\!\!\!\!\!\!\!\!\Gamma(P_{\rm B}b_{k,\uparrow}^\dag,P_{\rm B}b_{-k,\downarrow}^\dag|\omega)&=&{\rm i}u(\epsilon_k)^*v(\epsilon_k)^*\left(\frac{n(\epsilon_k)}{\omega+\omega(\epsilon_k)+{\rm i}\eta}-\frac{1-n(\epsilon_k)}{\omega-\omega(\epsilon_k)+{\rm i}\eta}\right).
\end{eqnarray}
At the end of the calculation we shall take the limit $\eta\rightarrow 0^+$. Further, we assume an energy  and spin independent coupling to environments $\kappa=\pi\sum_k|t_{k,\sigma}|^2\delta(\epsilon-\epsilon_k)$ and obtain
\begin{eqnarray}\nonumber
\!\!\!\!\!\!\!\!\!\!\!\!\!\!\!\!\!\!\!\!\!\!\!\!\!\!\!\!\!\!\!\!\!\!\!\!\!\gamma^{(1)}(\omega,\eta)&=&\sum_{k}|t_{k,\sigma}|^2\gamma(P_{\rm B}b_{k,\sigma}^\dag,b_{k,\sigma}P_{\rm B}|\omega)\\ \nonumber
\!\!\!\!\!\!\!\!\!\!\!\!\!\!\!\!\!\!\!\!\!\!\!\!\!\!\!\!\!\!\!\!\!\!\!\!\! &=&\sum_k \int_{-\infty}^\infty{\rm d}\epsilon\,\delta(\epsilon-\epsilon_k)|t_{k,\sigma}|^2\left(\frac{\eta  |u(\epsilon_k)|^2 n_k}{\eta ^2+\left(\omega _k+\omega \right){}^2}+\frac{\eta  |v(\epsilon_k)|^2 \left(1-n_k\right)}{\eta^2+\left(\omega -\omega _k\right){}^2}\right)\\ 
\!\!\!\!\!\!\!\!\!\!\!\!\!\!\!\!\!\!\!\!\!\!\!\!\!\!\!\!\!\!\!\!\!\!\!\!\!&\approx&\frac{\kappa}{\pi} \int_{-\infty}^\infty{\rm d}\epsilon\,\eta\left(\frac{ |u(\epsilon)|^2 n(\epsilon)}{\eta ^2+\left(\omega(\epsilon)+\omega \right){}^2}+\frac{ |v(\epsilon)|^2 \left(1-n(\epsilon)\right)}{\eta^2+\left(\omega -\omega(\epsilon)\right){}^2}\right)\\ \nonumber
\!\!\!\!\!\!\!\!\!\!\!\!\!\!\!\!\!\!\!\!\!\!\!\!\!\!\!\!\!\!\!\!\!\!\!\!\! &=&\frac{\kappa}{\pi} \int_{0}^\infty{\rm d}\epsilon\,\eta\left(\frac{n(\epsilon)}{\eta ^2+\left(\omega(\epsilon)+\omega \right){}^2}+\frac{1-n(\epsilon)}{\eta^2+\left(\omega -\omega(\epsilon)\right){}^2}\right),\\ \nonumber
\!\!\!\!\!\!\!\!\!\!\!\!\!\!\!\!\!\!\!\!\!\!\!\!\!\!\!\!\!\!\!\!\!\!\!\!\! \gamma^{(1)}(\omega)&=&\lim_{\eta\rightarrow0}\gamma^{(1)}(\omega,\eta)=\kappa\rho_L(\omega)(1-n(\omega)),\\ \nonumber
\!\!\!\!\!\!\!\!\!\!\!\!\!\!\!\!\!\!\!\!\!\!\!\!\!\!\!\!\!\!\!\!\!\!\!\!\! \gamma^{(2)}(\omega,\eta)&=&\sum_{k}|t_{k,\sigma}|^2\gamma(b_{k,\uparrow}P_{\rm B},b_{-k,\downarrow}P_{\rm B}|\omega)\\ \nonumber
\!\!\!\!\!\!\!\!\!\!\!\!\!\!\!\!\!\!\!\!\!\!\!\!\!\!\!\!\!\!\!\!\!\!\!\!\! &\approx&\frac{\kappa}{\pi} \int_{0}^\infty{\rm d}\epsilon \,u(\epsilon)v(\epsilon)\eta\left(\frac{n(\epsilon)}{\eta ^2+\left(\omega(\epsilon)+\omega \right){}^2}-\frac{1-n(\epsilon)}{\eta^2+\left(\omega -\omega(\epsilon)\right){}^2}\right),\\ \nonumber
\!\!\!\!\!\!\!\!\!\!\!\!\!\!\!\!\!\!\!\!\!\!\!\!\!\!\!\!\!\!\!\!\!\!\!\!\! \gamma^{(2)}(\omega)&=&\lim_{\eta\rightarrow0}\gamma^{(2)}(\omega,\eta)=2\kappa\rho_L(\omega)(1-n(\omega)) u(\omega)v(\omega)(\Theta(\omega)-\Theta(-\omega)),
\end{eqnarray}
where $\rho_L(\omega)=\Theta(\omega-\Delta)\omega/\sqrt{\omega^2-\Delta^2}$ is the superconducting density of states and $\Theta(\omega)$ is the Heaviside step function. All other correlation functions are up to a sign equal to $\gamma^{(1)}(\omega)$ or $\gamma^{(2)}(\omega)$. In order to determine the Lamb shift we have to calculate the following sums
\begin{eqnarray}
\label{eq:corr_s}
\!\!\!\!\!\!\!\!\!\!\!\!\!\!\!\!\!\!\!\!\!\!\!\!\!\!\!\!\!\!\!\!S^{(1)}(\omega)=\sum_{k}|t_{k,\sigma}|^2S(P_{\rm B}b_{k,\sigma}^\dag,b_{k,\sigma}P_{\rm B}|\omega)=\sum_{k}|t_{k,\sigma}|^2S(b_{k,\sigma}P_{\rm B},P_{\rm B}b_{k,\sigma}^\dag|\omega),\\ \nonumber
\!\!\!\!\!\!\!\!\!\!\!\!\!\!\!\!\!\!\!\!\!\!\!\!\!\!\!\!\!\!\!\!S^{(2)}(\omega)=\sum_{k}|t_{k,\sigma}|^2S(b_{k,\uparrow}P_{\rm B},b_{-k,\downarrow}P_{\rm B}|\omega),\\ \nonumber
\!\!\!\!\!\!\!\!\!\!\!\!\!\!\!\!\!\!\!\!\!\!\!\!\!\!\!\!\!\!\!\!\sum_{k}|t_{k,\sigma}|^2S(b_{k,\downarrow}P_{\rm B},b_{-k,\uparrow}P_{\rm B}|\omega)=-S^{(2)}(\omega),\\ \nonumber
\!\!\!\!\!\!\!\!\!\!\!\!\!\!\!\!\!\!\!\!\!\!\!\!\!\!\!\!\!\!\!\!\sum_{k}|t_{k,\sigma}|^2S(P_{\rm B}b_{k,\uparrow}^\dag,P_{\rm B}b_{-k,\downarrow}^\dag|\omega)=-S^{(2)*}(\omega),\\ \nonumber
\!\!\!\!\!\!\!\!\!\!\!\!\!\!\!\!\!\!\!\!\!\!\!\!\!\!\!\!\!\!\!\!\sum_{k}|t_{k,\sigma}|^2S(P_{\rm B}b_{k,\downarrow}^\dag,P_{\rm B}b_{-k,\uparrow}^\dag|\omega)=S^{(2)*}(\omega).
\end{eqnarray}
This can be done numerically using the relations (\ref{eq:ap_gamma}) and the definitions (\ref{eq:corr_def}). The results are independent of the used bandwidth in the sums (\ref{eq:corr_s}).

\end{document}